\documentclass{jfm}

\usepackage{enumerate}
\usepackage{color}
\usepackage{graphicx}
\usepackage{natbib}
\usepackage{bm}
\usepackage{amsmath}
\ifCUPmtlplainloaded \else
  \checkfont{eurm10}
  \iffontfound
    \IfFileExists{upmath.sty}
      {\typeout{^^JFound AMS Euler Roman fonts on the system,
                   using the 'upmath' package.^^J}%
       \usepackage{upmath}}
      {\typeout{^^JFound AMS Euler Roman fonts on the system, but you
                   dont seem to have the}%
       \typeout{'upmath' package installed. JFM.cls can take advantage
                 of these fonts,^^Jif you use 'upmath' package.^^J}%
      }
  \else
  \fi
\fi

\ifCUPmtlplainloaded \else
  \checkfont{msam10}
  \iffontfound
    \IfFileExists{amssymb.sty}
      {\typeout{^^JFound AMS Symbol fonts on the system, using the
                'amssymb' package.^^J}%
       \usepackage{amssymb}%

      }{}
  \fi
\fi

\ifCUPmtlplainloaded \else
  \IfFileExists{amsbsy.sty}
    {\typeout{^^JFound the 'amsbsy' package on the system, using it.^^J}%
     \usepackage{amsbsy}}
    {}
\fi

\newsavebox{\astrutbox}
\sbox{\astrutbox}{\rule[-5pt]{0pt}{20pt}}

\title[Flux correlations in supersonic turbulence]
{Flux Correlations in Supersonic Isothermal Turbulence}

\author[R. Wagner, G. Falkovich, A. G. Kritsuk and M. L. Norman]%
{R.\ns W\ls A\ls G\ls N\ls E\ls R$^1$%
  \thanks{Email address for correspondence: rpwagner@sdsc.edu},\ns
G.\ns F\ls A\ls L\ls K\ls O\ls V\ls I\ls C\ls H$^2$\break A.\ls
G.\ns K\ls R\ls I\ls T\ls S\ls U\ls K\ls$^3$ \and M.\ns L.\ns N\ls
O\ls R\ls M\ls A\ls N$^{1,3}$}

\affiliation{$^1$San Diego Supercomputer Center, University of California,
San Diego, MC 0505, 10100 Hopkins Drive, La Jolla, CA 92093-0505, USA\\[\affilskip]
$^2$Physics of Complex Systems, Weizmann Institute of Science, Rehovot 76100, Israel\\[\affilskip]
$^3$Department of Physics and Center for Astrophysics and Space
Sciences, University of California, San Diego, MC 0424, 9500 Gilman
Drive, La Jolla, CA 92093-0424, USA}

\pubyear{2012}
\volume{7XX}
\pagerange{NNN--MMM}
\date{?; revised ?; accepted ?. - To be entered by editorial office}

\begin{document}

\maketitle

\begin{abstract}
Using data from a large-scale three-dimensional simulation of
supersonic isothermal turbulence, we have tested the validity of an exact
flux relation derived analytically from the Navier--Stokes equation by
Falkovich, Fouxon and Oz [2010 {New relations for correlation functions in Navier--Stokes turbulence}.
{\em J. Fluid  Mech.} {\bf 644}, 465]. That relation, for compressible barotropic fluids,
was derived assuming turbulence generated by a large-scale force. However,
compressible turbulence in simulations is usually initialized and maintained by a
large-scale acceleration, as in gravity-driven astrophysical flows.
We present a new approximate flux relation
for isothermal turbulence driven by a large-scale acceleration, and
find it in reasonable agreement with the simulation results.
\end{abstract}

\section{Introduction}

An important rigorously derived result in the statistical theory of incompressible turbulence is
Kolmogorov's four-fifths law~\citep[]{Kolmogorov1941c}
\begin{equation}
  \left\langle[\delta u_{\parallel}(r)]^3\right\rangle = -\frac{4}{5}\epsilon r,
\label{four-fifths}
\end{equation}
where
\begin{equation}
  \delta u_{\parallel}(r)\equiv \left[\bm u(\bm r) - \bm u(0)\right]\cdot \bm r/r
\end{equation}
is the longitudinal velocity difference between two points, $0$
and $\bm r$, separated by a distance $r = |\bm r|$, and $\epsilon$ is the mean energy
dissipation rate. The four-fifths law is well supported experimentally and
has been traditionally interpreted as a signature of a direct kinetic energy
cascade within the inertial range of scales. Recent work by~\citet[][hereafter FFO]{Falkovich2010},
reinterprets the Kolmogorov relation in terms of currents and densities of the conserved
quantities. From this new perspective, FFO derived an analytic scaling
relation  for barotropic flows with an arbitrary degree of compressibility,
based on the correlation of currents and conserved fluxes in the inertial range.

There are few  analytic relations proposed for compressible turbulence,
e.g., the relation for an isothermal gas based on an effective energy transfer rate \citep{GaltierBanerjee2011}, and
the lack of experimental results at high turbulent Mach numbers makes verification of such relations
challenging. While supersonic turbulence is observed in molecular clouds~\citep{Heyer:2004fk},
and is believed to play an important role in star formation~\citep{McKee:2007uq}, observations
are limited by available angular resolution, projection and finite optical depth effects~\citep{ElmegreenScalo}.
At present, numerical experiments provide the best route for testing theories of compressible
turbulence~\citep{Sytine:2000fk,Benzi:2008fj,Kitsionas:2009fk,Kritsuk1511,Brandenburg11}.
Three-dimensional numerical simulations show that even at very high Mach numbers, a compressible
energy cascade can be recovered with a proper density weighting,
$\rho^{1/3}\bm u$~\citep{Kritsuk2007fk,K07b,Pan:2009lr}. \citet{Aluie2012}
have recently presented evidence from simulations for pressure-dilatation action
at large scales, and kinetic energy transfer to the dissipation scale via a conservative
cascade, supporting an earlier analytical proof of the locality of kinetic energy transfer in
compressible turbulence~\citep{Aluie2011}.

In this article, we use data from a simulation of supersonic isothermal
turbulence to analyze the exact relation for compressible fluids from
FFO. We are particularly interested in evaluating the role of the 
properties of the force that acts as a source of momentum and energy to compensate
for the dissipation losses. We discuss the limitations implied by the assumptions used in the
FFO derivation and present a new approximate scaling relation appropriate for the driving
commonly used in numerical experiments on compressible fluid turbulence
\citep[e.g.,][]{PorterPouquetWoodward,Kritsuk2007fk,Schmidt08,Wang10}.

\section{Flux Correlations}
\label{fluxcorr_sec}

A general relation is derived in FFO, along with a particular case
for compressible turbulence in a barotropic fluid, both of which are
based on two-point statistics in the inertial range. The
lower limit of $r$ where the relations are expected to hold
is the scale where dissipation is
negligible, while the upper limit is the correlation length of
the force. The correlations involve the
densities $q^a$, currents $\bm j^a$, fluxes $F_i^a$, and sources
(forcing) $f^a$; the sources are assumed to be random, statistically
stationary, spatially homogeneous, and isotropic. Beginning with the
governing equations, and eliminating terms tied to dissipation, a
relation involving the densities, fluxes, and forcing is derived:
\begin{equation}
\nabla_i\left\langle q^a(0, t)F_i^a(\bm r, t)\right\rangle = \left\langle q^a(0, t) f^a(\bm r, t)\right\rangle,
\label{flux_corr_scale_dependent}
\end{equation}
see FFO for more detail.

In order to eliminate the scale dependence on the right-hand side of
(\ref{flux_corr_scale_dependent}), the authors assume that on scales
much smaller than the correlation length ($r\ll L$),
the force is approximately constant,
\begin{equation}
f^a(0, t)\approx f^a(\bm r, t).
\label{f_approx}
\end{equation}
Therefore, in the inertial range, the correlation of a density and its
source is also approximately constant,
\begin{equation}
  \left\langle q^a(0, t) f^a(\bm r, t)\right\rangle \approx
  \left\langle q^a(0, t) f^a(0, t)\right\rangle\equiv \bar\epsilon_a.
  \label{force_assumption}
\end{equation}
Note that this ansatz is essentially carried over from the incompressible case, where there is no
distinction between the large-scale acceleration and large-scale force since the density is constant.%
\footnote{While we recognize that the assumption of large-scale force per unit volume is possibly the
only viable path to a rigorously derived relation \citep[][also relied on the same constraint to come up
with an alternative relation for a compressible case]{GaltierBanerjee2011}, we emphasize that
driving variable-density flows with a large-scale force would have a nontrivial effect on the turbulence
statistics and eliminate the inertial range in the traditional sense, with respect to the fluid velocity.}
Substituting $\bar\epsilon_a$ into (\ref{flux_corr_scale_dependent})
gives
the correlation function
\begin{equation}
  \nabla_i\left\langle q^a(0)F_i^a(\bm r)\right\rangle = \bar\epsilon_a\label{densfluxcorr_grad},
\end{equation}
and assuming isotropy, FFO derive the exact scaling relation in a vector form:
\begin{equation}
  \left\langle q^a(0)F_i^a(\bm r)\right\rangle = \frac{\bar\epsilon_a r_i}{d}\label{densfluxcorr},
\end{equation}
where $d$ is the number of spatial dimensions.

In the case of a barotropic fluid, the densities are $\bm q = (\rho \bm u,
\rho)$, and the fluxes are $F_i^j = \rho u_i u_j +
p(\rho)\delta_{ij}$ for $i,j=1,\,\ldots,\,d$, while $F_i^{d+1} = \rho u_i$. When dissipation is neglected, the governing equations are the
forced Euler equations:
\begin{eqnarray}
\partial_t\rho+\bm{\nabla}\cdot(\rho\; \bm{u}) & = & 0,\label{euler1}\\
\partial_t(\rho u_i)+ \partial_j(\rho u_i u_j + p\delta_{ij}) & =& f^i\label{euler2}.
\end{eqnarray}
Substituting the densities, fluxes, and forcing into (\ref{densfluxcorr}) gives a relation for
isothermal turbulence when $p = \rho$,
\begin{equation}
\Phi(r)\equiv\left\langle\left[\rho(0)\bm u(0)\cdot\bm u(\bm r)\rho(\bm r)\right]
  u_\parallel(\bm r)\right\rangle + \left\langle\rho(0)  u_\parallel(0)\rho(\bm r)\right\rangle = \frac{\bar \epsilon r}{d},
  \label{nsrel}
\end{equation}
where $u_\parallel = \bm u\cdot\bm r/r$ is the longitudinal velocity,
\begin{equation}
  \bar{\epsilon} = \left\langle\rho(0)\bm{u}(0)\cdot\bm{f}(0)\right\rangle,
  \label{epsilonbar}
\end{equation}
and we have taken the scalar product of both sides of (\ref{densfluxcorr})
with $\bm r/r$ to get (\ref{nsrel}) in a more convenient scalar form.
Unlike $\epsilon$ in (\ref{four-fifths}), $\bar{\epsilon}$ is the
injection rate of momentum squared, not energy. However, in the
incompressible limit, the vector form of (\ref{nsrel}) reduces to a
third-order velocity correlation function implying (\ref{four-fifths}).

In numerical experiments of compressible turbulence, the flow is traditionally
driven by a large-scale acceleration $\bm a$, such that $\bm f = \rho\bm a$, while the exact
 flux relation (\ref{nsrel}) was derived under the assumption that
 ${\bm f}$ is a smooth, large-scale, field. As noted in FFO,
 physically, $\bm f$ may come from a gradient of potential (e.g.,
 external gravitational acceleration, i.e. $\bm f = \rho\nabla\phi$ or $\bm a=\nabla\phi$).
 The decorrelation and small-scale variations of the density at
 high Mach number make this an important point, as acceleration-driven
 turbulence is important in astrophysics and other areas, such as turbulent convection.

Let us suggest a new flux relation appropriate for this case. Now,
instead of the constant $\bar\epsilon$ in  (\ref{epsilonbar}) we shall
have the two-point fourth-order correlation function
$\langle\rho(0)\rho(\bm r){\bm u}(0)\cdot{\bm a(\bm r)}\rangle$, which
is scale-dependent. The character of this scale-dependence can be
found in the particular case of an acceleration short-correlated in
time, when $\langle {a }_i(t){a}_j(0)\rangle
=\varepsilon_{ij}\delta(t)$. The average of the product of any
quantity $U\{{\bm a }\}$  and a white noise ${\bm a }$ is expressed by
the formula of Gaussian integration via a variational derivative:
$\langle U\{{\bm a }\}a_i\rangle=\varepsilon_{ij}\langle\delta U/\delta a_j\rangle$.
In the correlation function we consider, it is
the velocity field which is related to the acceleration (by the
equation $du_i/dt-u_i\nabla\cdot{\bm u}=a_i$) so that $\langle\delta
u_i(t)/\delta a_j(t')\rangle=\delta_{ij}\theta(t-t')$ where the last
factor is the step function. We now obtain
\begin{equation}
\left\langle\rho(0)\rho(\bm r){\bm u}(0)\cdot{\bm a}(\bm
  r)\right\rangle=\left\langle\rho(0)\rho(\bm
  r)\right\rangle\bar\varepsilon\,
\label{short}
\end{equation}
where $\bar\varepsilon=\langle {\bm u}(0)\cdot{\bm  a}(0)\rangle$.
Since the acceleration is a large-scale field, then a
two-point velocity-acceleration correlation function can be replaced
by a single-point one. Let us stress that the constant $\bar\varepsilon$
is neither the total energy input nor the work of the external acceleration.

Of course, in the physical situations of interest, as well as in
simulations, the acceleration cannot be considered short-correlated
in time. In this case, the effective decoupling of the fourth moment
into the product of second moments expressed by (\ref{short}) is
not an exact relation  but can be suggested as a plausible approximation.
We thus come to the generalization of the flux relation in the following
form (up to an order-unity constant):
\begin{equation}
\nabla_r\Phi(r)\simeq\langle\rho(0)\rho(\bm r)\rangle
\bar\varepsilon\label{gen1}.
\end{equation}
For isothermal turbulence driven by a large-scale acceleration, (\ref{gen1})
provides an approximate relation connecting the scaling of $\Phi(r)$
with that of $\left\langle\rho(0)\rho(\bm r)\right\rangle$ in the intertial range.

\section{Experimental Verification}
To appraise the relations from FFO, we have used data from a simulation designed to study
the inertial range statistics of highly compressible turbulence~\citep{Kritsuk2007fk}.
The simulation was performed with an implementation of the piecewise parabolic
method~\citep{PPM} in the Enzo code~\citep{PetaEnzo}. The forced Euler equations
 for an isothermal gas, (\ref{euler1}) and (\ref{euler2}), were solved numerically in
a cubic periodic domain with a linear size of $L = 1$ along each axis, and covered by a uniform Cartesian grid of
$1024^3$ zones, each having sides of length $\Delta=L/1024$. The turbulent rms
Mach number, $M = 6$, and a steady rate of energy injection were maintained by a
random acceleration field with power limited to wavenumbers $k/k_{min} \in
[1, 2]$, where $k_{min} = 2\pi/L$. The spatially fixed acceleration field was
normalized at every time step so that $\langle\rho\bm u\cdot\bm
a\rangle$ was constant; the normalization factor had a standard
deviation of $\sim 5\%$ during the simulation. For this work, we used
$43$ full data snapshots
evenly distributed in time in the range $t/\tau \in[6,10]$, where the flow crossing
(large eddy turnover) time,  $\tau\equiv L/2M$, assumes
the sound speed of unity. For each snapshot, we computed $\bar\epsilon$, $\bar\varepsilon$,
and evaluated $\Phi(r)$ for 16 discrete $r$-values in the range $r/\Delta \in[8, 128]$, using $2^{32}
\approx 4\times10^9$ random point pairs for each value of $r$.

\begin{figure}
  \centering
  \includegraphics[width=4.55in]{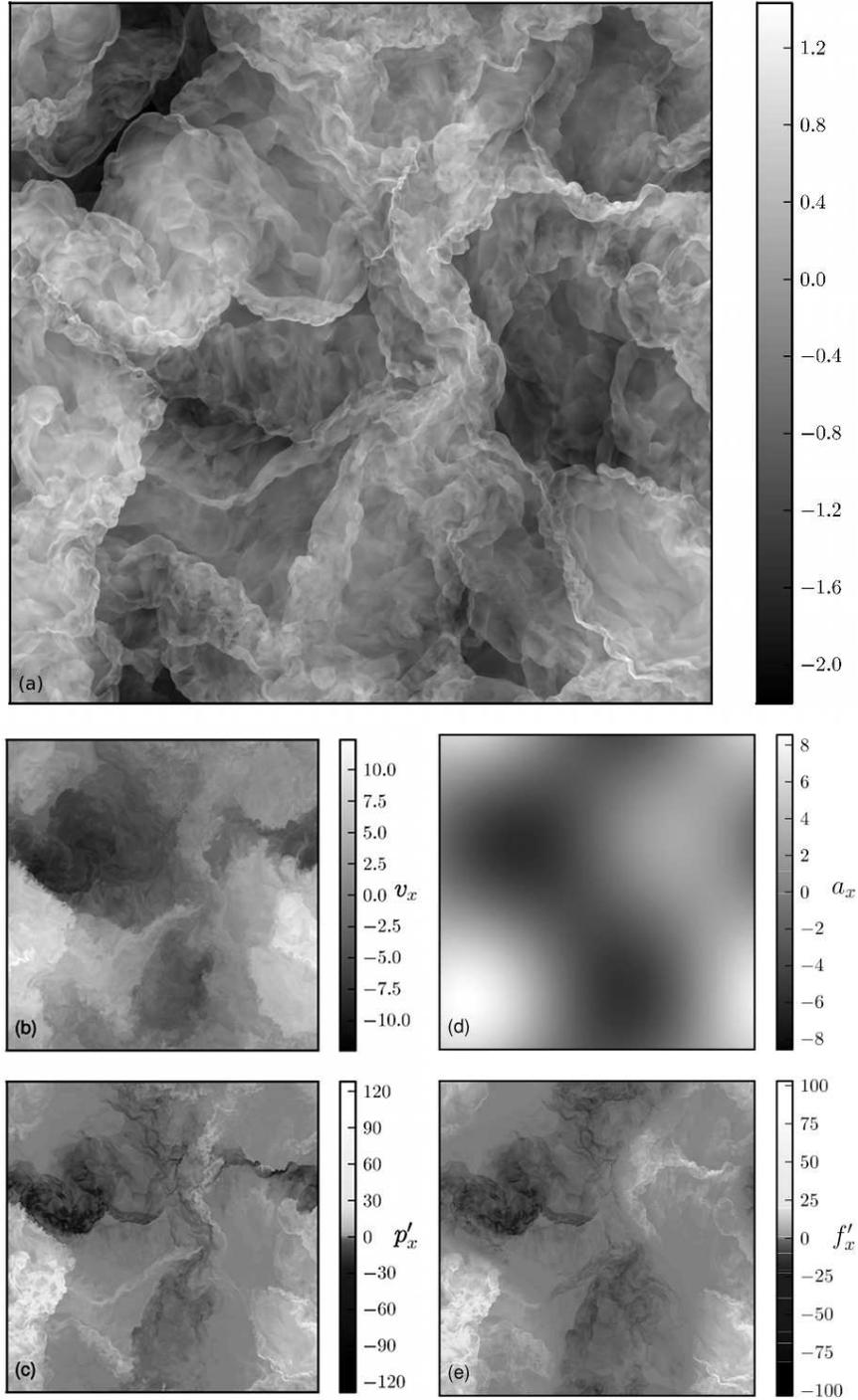}
  \caption{Slices of several instantaneous dynamical fields through
    the plane $x = 0$: (a) logarithm of the normalized density field $\log_{10}(\rho/\bar{\rho})$
    ; (b) velocity field $u_x$ (velocity is normal to
    the image plane); (c) normalized momentum $p'_x =
    \rho/\bar{\rho}\ u_x$; (d) acceleration field $a_x$; (e) normalized force $f'_x =
    \rho/\bar{\rho}\ a_x$. N.B.: To account for the distribution of the
    density field, the colour maps used for $p'_x$ and $f'_x$ are highly
    compressed around $0$.}
  \label{axx_plot}
\end{figure}

To evaluate the role of various assumptions concerning the forcing and assess the hypotheses
employed by FFO, we compare $\Phi$ based on the exact forms (\ref{densfluxcorr_grad}) and
(\ref{nsrel}) with an approximate expression based on (\ref{gen1}). We begin by examining
the spatial distribution of the force and acceleration in the lower right-hand panels of
figure~\ref{axx_plot}, which show the values of the $x$ component of $\bm a$ and $\bm f'$, in
a slice through the simulation volume at $x = 0$, where $\bm f' \equiv
\rho/\bar{\rho}\bm a$ is the force normalized by the average density.
\begin{figure}
  \centering
  \includegraphics[width=\columnwidth]{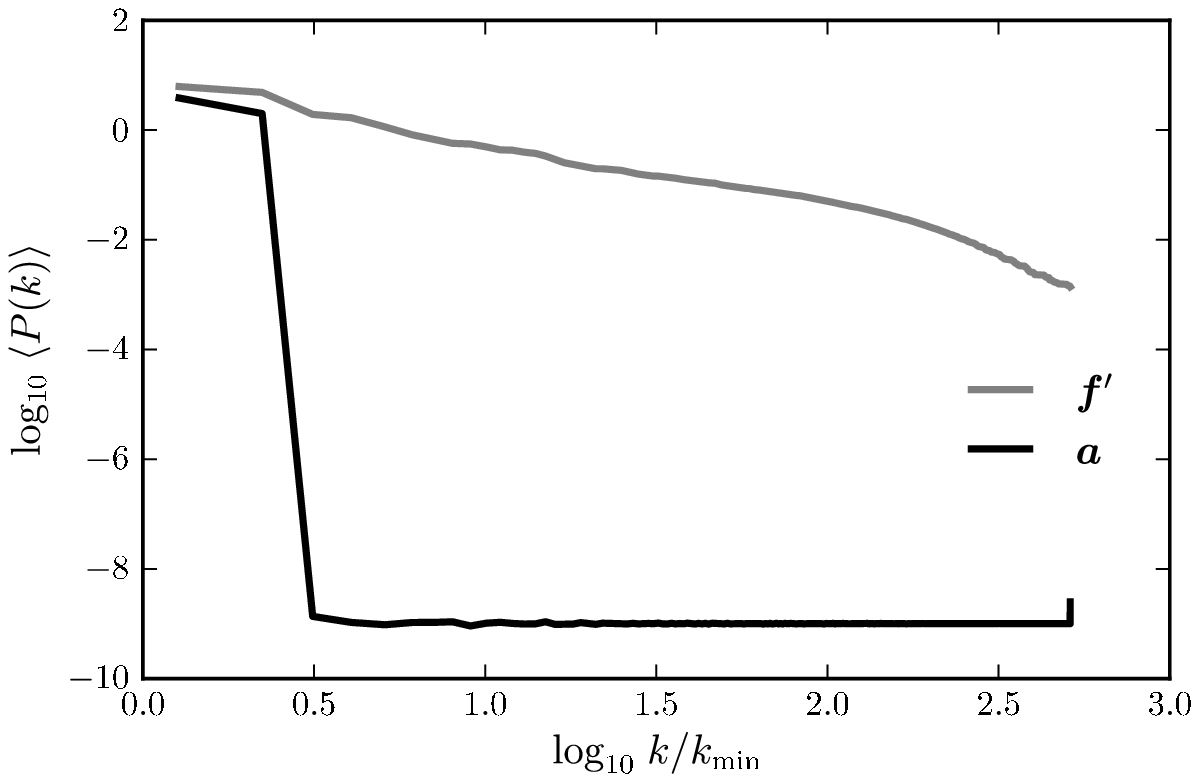}
  \caption{Power spectra of the driving acceleration,
    $\bm a$ (solid black line), and the normalized force, $\bm
    f' = \rho\bm a/\bar{\rho}$ (solid grey
    line), from a single flow snapshot.}
  \label{force_power}
\end{figure}
A more quantitative assessment can be found in figure~\ref{force_power}, which
shows the power spectra of both $\bm a$ and $\bm f'$. The power spectrum of
$\bm  a$ falls off at $k/k_{min} > 2$, while the spectrum of $\bm f'$
slowly decreases with $k$, and the two are separated by several orders
of magnitude above $k/k_{min} \approx 2$. Further, we measure $\bar\epsilon = 589$,
$\bar\varepsilon = 149$,  $\left\langle\rho^2\right\rangle = 3.74$,
and find that
\begin{equation}
\left\langle\rho^2\right\rangle\bar\varepsilon = 0.95\ \bar\epsilon\label{decoup},
\end{equation}
supporting the decoupling suggested in (\ref{short}). This decoupling, combined
with the features of $\bm f$ shown by figures~\ref{axx_plot} and~\ref{force_power},
indicates that the assumption of an approximately constant force does not
hold in the simulation, as expected.

Next, we will compare $\Phi$ to the scaling from (\ref{gen1}).
\begin{figure}
\includegraphics[width=\columnwidth]{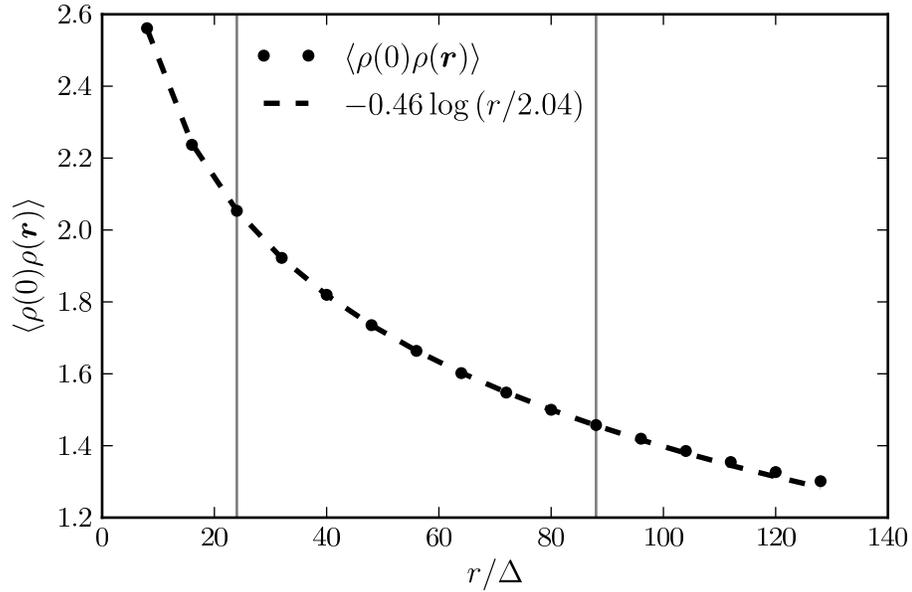}
\caption{Dots: time-averaged density correlation;  the standard error of the mean of $\left\langle\rho(0)\rho(\bm
    r)\right\rangle$ is $< 0.02$. Dashed line: best fit to the data points with $a\log(r/b)$
    in the range $r/\Delta \in[20, 90]$, as indicated by the vertical grey lines.}
\label{denscorr}
\end{figure}
We find that the density correlation is well approximated by a logarithmic form:
\begin{equation}
\left\langle\rho(0)\rho(\bm
  r)\right\rangle \simeq a\log\left(\frac{r}{b}\right)\label{denslog}
\end{equation}
with $a = -0.46 \pm 0.001$ and $b = 2.04  \pm 0.01$
when fitted in the range $r/\Delta \in[20, 90]$ (see
figure~\ref{denscorr}). This is expected from the density
power spectrum approximately following $\sim k^{-1}$ at this particular turbulent
Mach number~\citep{Kritsuk2007fk}. Substituting (\ref{denslog}) into
(\ref{gen1}) and integrating radially from $0$ to $r$ assuming
isotropy, we have
\begin{equation}
\Phi(r)  \simeq \frac{a\bar\varepsilon}{9}\left[3\log\left(\frac{r}{b}\right) -
1\right]r
\label{phi_scale_fit_log}.
\end{equation}
Note that (\ref{phi_scale_fit_log}) is a good approximation only for turbulence
at $M = 6$. Simulations with different Mach numbers will have density power
spectra with different slopes, and their density correlations will have different
forms.

\begin{figure}
  \centering
  \includegraphics[width=\columnwidth]{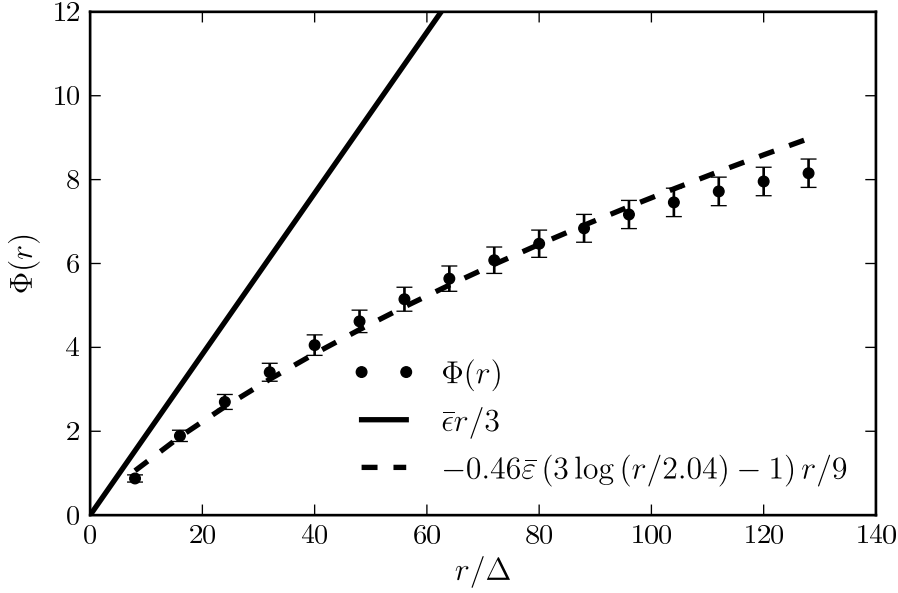}
  \caption{Dots: time-averaged measure of
    $\Phi(r)$; the error bars show the standard error of the
    mean. Solid line: expected scaling based on a large-scale
    force. Dashed line: expected approximate scaling based on the
    density correlation. Note that the dashed line is not a fit to the
    $\Phi(r)$ points, but shows the right-hand side of (\ref{phi_scale_fit_log}) with
    parameters determined by the best fit to the density correlation
    (\ref{denslog}) shown in figure~\ref{denscorr}.}
  \label{phi_plot}
\end{figure}

Figure~\ref{phi_plot} shows the time-averaged $\Phi(r)$, the expected
scaling for a large-scale force using (\ref{nsrel}), and the
expected scaling based on a large-scale acceleration and our
approximation  (\ref{phi_scale_fit_log}).
\begin{figure}
  \centering
  \includegraphics[width=\columnwidth]{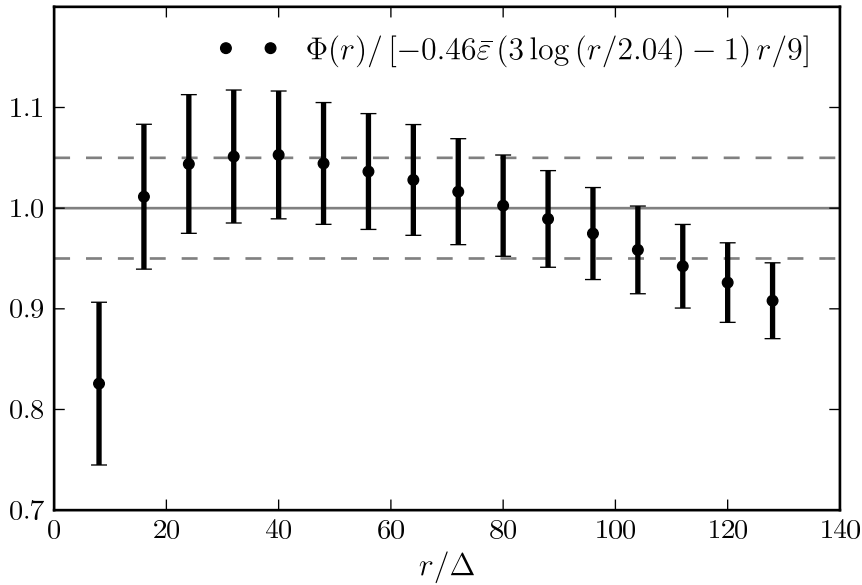}
  \caption{Time-averaged measure of $\Phi(r)$ normalized by the
    expected approximate scaling~(\ref{phi_scale_fit_log}). The dashed
    grey lines show $\pm5$\% tolerance levels.}
  \label{phi_comp}
\end{figure}
The accuracy of the approximate flux relation can be assessed in
figure~\ref{phi_comp}, where we show $\Phi(r)$ normalized by the
right-hand side of (\ref{phi_scale_fit_log}), using the values fitted to
the density correlation. The right-hand side of (\ref{phi_scale_fit_log}) follows
the actual measured $\Phi(r)$ within $\pm 5\%$ in the range $r/\Delta \in[10, 100]$,
somewhat underestimating $\Phi$ at the lower end of the range, and overestimating it as $r$
increases; this behaviour can be predicted as the assumed approximations
break down for two different reasons.
First, from (\ref{decoup}), the density correlation and injection rate
of specific kinetic energy are not totally decoupled; this may account
for (\ref{gen1}) underestimating $\Phi$ at small scales. Second, as $r$ approaches $L/2$, $\langle {\bm
  u}(0)\cdot{\bm a}(\bm r)\rangle \to 0$, and as (\ref{f_approx}) breaks
down, a constant $\bar\varepsilon$ on the right-hand side of
(\ref{gen1}) causes it to excede $\Phi(r)$. This departure by $\Phi(r)$
from (\ref{phi_scale_fit_log}) at large scales is determined by the
numerical resolution and driving scales; given a similar simulation with
higher numerical resolution, we would expect the agreement to scale
relative to the smallest driving scale.

\section{Conclusion}
We find that the approximate relation we derived for supersonic
isothermal turbulence driven by a large-scale acceleration field
(\ref{gen1}) agrees reasonably well with the simulation results within
a sufficiently wide range of scales consistent with the inertial range
presence established in \citet{Kritsuk2007fk}. This implies
support for the original FFO exact relation
(\ref{flux_corr_scale_dependent}), so it can be anticipated that in a
system where (\ref{force_assumption}) holds, the exact relation
(\ref{nsrel}) will be satisfied as well. Therefore, figures 4 and 5
demonstrate the strong sensitivity of the flux correlations in
supersonic isothermal turbulence to the nature of forcing supporting
the statistically stationary state. Such sensitivity has been observed
in a similar context of the forced one-dimensional Burgers equation, where
an invasive random force with a $\propto k^{-1}$ spectrum (compare with the grey line
in figure~\ref{force_power}) would produce
a non-universal bifractal scaling \citep{mitra...05}. Likewise, \citet{bahraminasab.....08}
showed that the same Burgers system driven by a force with
large-scale correlation in space and with a Wiener scaling in time exhibits a
non-universal scaling for time increments larger than the correlation time of the force.
Comparison with other relations and results
addressing energy transfer in compressible turbulence
~\citep{GaltierBanerjee2011,Aluie2011,Aluie2012} will help us to
understand whether non-universality, observed here due to the driving mechanism, 
is a more general property of highly compressible flows.

\section*{Acknowledgements}
R.W. is supported in part by CyberInfrastructure Research,
Education and Development at the San Diego Supercomputer Center
(SDSC); A.K. is supported in part by NSF grants AST-0908740 and
AST-1109570.  The simulation utilized TeraGrid computer time allocations
MCA98N020 and MCA07S014 at SDSC.
The analysis was performed on the Extreme Science and
Engineering Discovery Environment (XSEDE) resource Trestles at SDSC
under a Director's Discretionary Allocation. A.K. and G.F. were supported in
part by the Project of Knowledge Innovation Program (PKIP) of Chinese Academy of Sciences,
Grant No. KJCX2.YW.W10.

\bibliographystyle{jfm}

\bibliography{ns-corr-jfm-1}

\end{document}